\documentclass{ws-procs9x6}
\usepackage{epsfig}
\begin{document}
\newcommand{\beq}{\begin{equation}}
\newcommand{\beqa}{\begin{eqnarray}}
\newcommand{\beqn}{\begin{eqnarray}}
\newcommand{\eeqn}{\end{eqnarray}}
\newcommand{\eeqa}{\end{eqnarray}}
\newcommand{\eeq}{\end{equation}}

\title{Quantum Mechanics, Path Integrals and Option Pricing: \\
 Reducing the Complexity of Finance}
\author{Belal E. Baaquie$^{(1)}$, Claudio Corian\`{o}$^{(2)}$}
\author{and}
\author{ Marakani Srikant$^{(1)}$ }
\address{$^{(1)}$ 
National University of Singapore \\
Singapore 119260\\
phybeb@nus.edu.sg\\
srikant@srikant.org}
\address{$^{(2)}$ Dipartimento di Fisica\\ 
Universita' di Lecce \\
INFN Sezione di Lecce \\
Via Arnesano 73100, Lecce, Italy \\
claudio.coriano@le.infn.it }
\maketitle

\abstracts{ Quantum Finance represents the synthesis of the techniques
  of quantum theory (quantum mechanics and quantum field theory) to
  theoretical and applied finance. After a brief overview of the
  connection between these fields, we illustrate some of the methods
  of lattice simulations of path integrals for the pricing of options.
  The ideas are sketched out for simple models, such as the
  Black-Scholes model, where analytical and numerical results are
  compared. Application of the method to nonlinear systems is also
  briefly overviewed.  More general models, for exotic or
  path-dependent options are discussed.  }

\section{Introduction}
\footnote{Prep n.: UNILE-CBR-02-03.  Talk presented by 
Claudio Corian\`{o} at the Intl. Workshop 
``Nonlinear Physics: Theory and Experiment II'', Gallipoli, Lecce, Italy, 
June 28 - July 4, 2002}
Financial markets have undergone a tremendous growth in the last
decades and in order to meet the need of customers new and complex
financial instruments have been developed. Risk assessment models and
the quantification of returns, given the huge amount of trading
involved worldwide, requires more sophisticated approaches than in the
past.  Nonlinearities play a key role in all of this, and, from this
side, the field is probably largely unexplored.

In general, intensive numerical simulations and fast algorithms are
needed to obtain useful results. Analytical results are limited,
except for simple models such as the Black-Sholes model and other
similar models.

Use of a path integral formulation has some advantages. First, it is
in close relation to the lagrangean description of diffusion
processes, second, it opens the way to the use of quantum mechanical
methods.

We will introduce the field at a non-expert level through a crash
course (next few sections). Then we will come to brefly illustrate
where nonlinearities appear and review some of the simplest equations
one can write down, generalizing the Black-Scholes model. We will then
proceed to discuss the path integral formulation of the Black-Scholes
and the model of {\it barrier options}, resorting to a lagrangian path
integral formulation. Strategies to solve the path integral are then
briefly presented, together with some results.  Detailed simulations,
algorithms of analysis and related applications will be presented
elsewhere \cite{BCM}. The review sections are standard knowledge in
the field, and are based on \cite{BEB}.

\section{A view on Theoretical Finance}
\subsection{ Simulating the Complexity of Finance}
The simulation of financial markets can be modeled, from a theoretical
viewpoint, according to two separate approaches: a {\it bottom up}
approach and (or) a {\it top down} approach.

For instance, the modeling of financial markets starting from
diffusion equations and adding a noise term to the evolution of a
function of a stochastic variable is a top down approach. This type of
description is, effectively, a statistical one.

A bottom up approach, instead, is the modeling of artificial markets
using complex data structures (agent based simulations) using general
updating rules to describe the collective state of the market. The
number of procedures implemented in the simulations can be quite
large, although the computational cost of the simulation becomes
forbidding as the size of each agent increases. Readers familiar with
{\it Sugarscape Models} and the computational strategies based on {\it
  Growing of Artificial Societies} \cite{epstein} have probably an
idea of the enormous potentialities of the field. However, one would
expect that the bottom up description should become comparable to the
top down description for a very large number of simulated agents.
 
The bottom up approach should also provide a better description of
{\it extreme events}, such as crashes, collectively conditioned
behaviour and market incompleteness, this approach being of purely
algorithmic nature. A top down approach is, therefore, a model of {\it
  reduced complexity} and follows a statistical description of the
dynamics of complex systems (for an introduction see \cite{rajesh}).

\subsection{Forward, Futures Contracts and Options}
Let the price at time t of a security be $S(t)$. A specific good can
be traded at time t at the price $S(t)$ between a buyer and a seller.
The seller (short position) agrees to sell the goods to the buyer
(long position) at some time T in the future at a price F(t,T) (the
contract price).  Notice that contract prices have a 2-time dependence
(actual time t and maturity time T).  Their difference $\tau=T-t$ is
usually called {\it time to maturity}.  Equivalently, the actual price
of the contract is determined by the prevailing actual prices and
interest rates and by the time to maturity.

Entering into a forward contract requires no money, and the value of
the contract for long position holders and strong position holders at
maturity T will be 
\beq (-1)^p\left(S(T) -F(t,T)\right) \eeq 
where $p=0$ for long positions and $p=1$ for short positions.  {\it
  Futures Contracts } are similar, except that the after the contract
is entered, any changes in the market value of the contract are
settled by the parties. Hence, the cashflows occur all the way to
expiry unlike in the case of the forward where only one cashflow
occurs. They are also highly regulated and involve a third party (a
{\it clearing house}). Forward, futures contracts and, as we will see,
{\it options } go under the name of {\it derivative products}, since
their contract price F(t, T) depend on the value of the underlying
security $S(T)$.

To complete this crash course on financial instruments we need to
define {\it options}.  Options are derivatives that can be written on
any security and have a more complicated payoff function than the
futures or forwards. For example, a call option gives the buyer (long
position) the right (but not the obligation) to buy or sell the
security at some predetermined strike-price at maturity. A {\it payoff
  function} is the precise form of the price. Path dependent options
are derivative products whose value depends on the actual path
followed by the underlying security up to maturity. In the case of
path-dependent options, since the payoff may not be directly linked to
an explicit right, they must be settled by cash. This is sometimes
true for futures and plain options as well as this is more efficient.
 
\section{Langevin Evolution}
In the top down description of theoretical finance, a security $S(t)$
follows a random walk described by a Ito-Weiner process (or Langevin
equation) as
\beq
\frac{d\,S(t)}{S(t)}=\phi d t + \sigma R(t) d t, 
\label{langevin}
\eeq
where R(t) is a Gaussian white noise with zero mean and uncorrelated
values at time t and $t'$ $\langle R(t) R(t')\rangle =\delta(t -t')$.
$\phi$ is the drift term or expected return, while $\sigma$ is a
constant factor multiplying the random source $R(t)$, termed {\it
  volatility}.

As a consequence of Ito calculus, differentials of functions of random
variables, say $f(S,t)$, do not satisfy Leibnitz's rule, and for a
Ito-Weiner process with drift (\ref{langevin}) one easily obtains for
the time derivative of $f(S,t)$
\beq
\frac{d f}{d t}={\partial{f}\over \partial{t}}+ \frac{1}{2}\sigma^2 S^2 
 {\partial^2 f\over \partial{S}^2 }
+ \phi S {\partial{f}\over \partial{S}} + \sigma S 
{\partial{f}\over \partial{S}}R.
\label{deriv}
\eeq
The Black-Scholes model is obtained by removing the randomness of the
stochastic process shown above by introducing a random process
correlated to (\ref{deriv}). This operation, termed {\it hedging},
allows to remove the dependence on the white noise function $R(t)$, by
constructing a {\it portfolio} $\Pi$, whose evolution is given by the
short-term risk free interest rate $r$ 
\beq 
\frac{d\Pi}{dt}= r \Pi
\label{portf}.
\eeq
A possibility is to choose $\Pi=f -{\partial{f}\over \partial{S}}S$. 
This is a portfolio 
in which the investor holds an option $f$ and short sells
\footnote{short selling of the stock should be possible} an amount of the 
underlying security S proportional to ${\partial{f}\over \partial{S}}$. 
A combination of (\ref{deriv}) and (\ref{portf}) yields the 
Black-Scholes equation
\beq
{\partial{f}\over \partial{t}}+ \frac{1}{2}\sigma^2 S^2 
 {\partial^2 f\over \partial{S}^2 }
+ r S {\partial{f}\over \partial{S}}= r f.
\label{bs}
\eeq
There are some assumptions underlying this result. 
We have assumed absence of arbitrage, constant spot rate $r$, 
continuous balance of the portfolio, no transaction costs and 
infinite divisibility of the stock.  

The quantum mechanical version of this equation is obtained by 
a change of variable $S=e^x$, with $x$ a real variable.   
This yields 
\beq
{\partial{f}\over \partial{t}}= H_{BS} f 
\eeq
with an Hamiltonian $H_{BS}$ given by 

\beq
 H_{BS}=- \frac{\sigma^2}{2}\frac{\partial^2}{\partial x^2}
+ \left(\frac{1}{2}\sigma^2 - r\right)\frac{\partial}{\partial x} + r.
\eeq
Notice that one can introduce a quantum mechanical formalism and interpret 
the option price as a ket $|f\rangle$ in the basis of $|x\rangle$, 
the underlying security price.  
Using Dirac notation, we can formally reinterpret  
$f(x,t)= \langle x|f(t)\rangle$, as a projection of an abstract quantum state 
$|f(t)\rangle$ on the chosen basis. 

In this notation, the evolution of the option price can be formally written 
as $|f,t\rangle=e^{t H}|f,0\rangle$, for an appropriate Hamiltonian H.  

In the presence of a stochastic volatility, the description is more
involved, but also more interesting.

In general, the description of these processes is driven by
two correlated white noise functions $R_1$ and $R_2$ 
\beqa
\frac{d V} {d t} &=&\lambda + \mu V + \zeta V^\alpha R_1\nonumber \\
\frac{d V} {d t} &=&\phi S + \sigma \sqrt{V} + \mu V + \zeta V^\alpha R_2
\label{coupled}
\eeqa
with $V=\sqrt{\sigma}$ and $\langle R_1(t) R_2(t') \rangle=1/\rho 
\,\,\delta(t - t')$, $\rho$ being the correlation parameter. 
However, since volatility is not traded 
in the market (the market is said to be incomplete), 
perfect hedging is not possible, and an additional term, 
the {\it market price of volatility risk} $\beta(S,V,t,r)$,  
is in this case introduced. $\beta$ can be modeled appropriately. 
In some models \cite{MG}, 
a redefinition of the drift term $\mu$ in 
(\ref{coupled}) in the evolution of the volatility is 
sufficient to hedge such more complex portfolios, 
which amounts to an implicit choice of $\beta(S,V,t,r)$.
 We just quote the result 
 for 
the evolution of an option price in the presence of stochastic volatility, which, in 
the Hamiltonian formulation are given by 
\beq
{\partial{f}\over \partial{t}}= H_{MG} f 
\eeq

with 
\beqa
H_{MG} &=&- \left(r - \frac{e^{y}}{2}\right)\frac{\partial}{\partial x }
 - \left( \lambda e^{-y} + \mu - \frac{\zeta^2}{2}e^{2 y (\alpha -1)}\right) 
\frac{\partial}{\partial y} \nonumber 
 - \frac{e^{y}}{2}\frac{\partial^2}{\partial x^2} \\
&& - \rho \zeta e^{y(\alpha - 1/2)}\frac{\partial^2}{\partial x \partial y} - 
\zeta^2 \frac{e^{2 y(\alpha -1)}}{2}\frac{ \partial^2}{\partial y^2} + r.
\eeqa
which is nonlinear in the variables $x=\log(S)$ and $y=\log(V)$. For general values 
of the parameters, the best way to obtain the pricing of the options in this model 
is by a simulation of the path integral.

\section{Monte Carlo Simulations of option pricing}
Simulations of the functional integral are rather straightforward and
we should omit any detail about them since they have been known {\it
  ever since} by the high energy physics community.  However, to reach
out to a less specialized audience, we will provide a simple
illustration of the method.
  
Once the model is given, one determines the underlying lagrangean. 
We assume a discretization of the time to maturity $\tau$ in intervals 
$\epsilon=\tau/N$, with N an arbitrary (large) integer.  

For instance,  for the Black Scholes model one gets the action 
\beq
S_{BS}= \epsilon \sum_{i=1}^N L_{BS}(i)
\eeq
with
\beq
L_{BS}(i)=-\frac{1}{2 \sigma^2}\left( \frac{x_i - x_{i-1}}{\epsilon}
 + r - \frac{\sigma^2}{2}\right)^2
\eeq
where we have introduced discretized positions $(x_i)$ 
for the variable $x=\log(S)$ which identifies the quantum mechanical state of the system. We will refer to it as to the stock price. The propagator for the 
stock price is given by the pricing kernel
\beqa
p_{BS}(x,x',\tau)&=&\int DX_{BS}e^{S_{BS}}\nonumber \\
&=&\langle x|e^{-\tau H_{BS}}| x'\rangle\nonumber \\
\eeqa
with 
\beq
\int DX_{BS}=\Pi_{t=0}^{\tau} \int_{-\infty}^{\infty}dx(t). 
\eeq
For barrier options there is an analogous procedure, except that now we need 
to introduce a generic potential $V(x)$ in the corresponding Hamiltonian 
\beq
H_V=-\frac{\sigma^2}{2}\frac{\partial^2}{\partial x^2} + 
\left( \frac{1}{2}\sigma^2 - V(x)\right)\frac{\partial}{\partial x} + V(x).
\eeq
The pricing kernel is the fundamental quantity to compute using the functional 
integral. Related attempts can be found in the literature \cite{NM}.

For this purposes, we have used a standard Metropolis algorithm. 
If thermalization is slow, it is possible to resort to use 
sequentially Metropolis updates and cluster updates. 
The latter is an update for the embedded Ising dynamics in the lattice 
variables $x_i/|x_i|$ 
(Swendsen-Wang, Wolff), and is included in for a faster generation of the 
thermalized paths of the stock price $x(t)$.
  
For processes involving a stochastic volatility $(y=\log(V))$ 
the expression of the path integral is more complicated and can be found 
in \cite{BM}. From now on we will just consider the case of a 
constant volatility.
 
If we denote by $g(x,K)$ the payoff function, with a strike price $K$, 
in this case the value of the option (its price) is given by the Feynman-Kac formula 
\beq
f(t,x)=\int_{-\infty}^{\infty}d\,x'\langle x|e^{-(T-t)H_{BS}}|x'\rangle g(x',K)
.
\label{FK}
\eeq
In actual simulations, it is convenient to compute directly 
the option price rather than the propagator itself. 
The simulation is done by taking the initial point x fixed, and letting the 
final point evolve according to its quantum dynamics. In this way a path 
$(x,x')$ is generated. After the first thermalization, $x'$ is allowed to 
undergo quantum fluctuations, at fixed x. Each $x'$ is then 
convoluted with the payoff function and an average is performed. Finally, this 
procedure is repeated for several x values, so to obtaint the option price 
at time to maturity $\tau$.
\begin{figure}[th]
\centerline{\includegraphics[angle=-90,width=1.\textwidth]{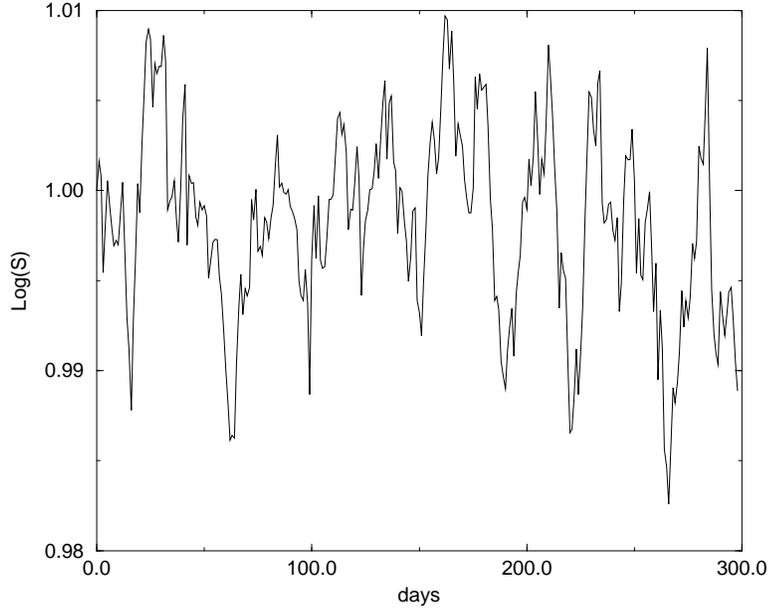}}
\caption{An example of thermalized path obtained from the simulation of the path integral (Black-Scholes) with r=0.05 and $\sigma=0.12$ }
\end{figure}
\begin{figure}[th]
\centerline{\includegraphics[angle=-90,width=1.2\textwidth]{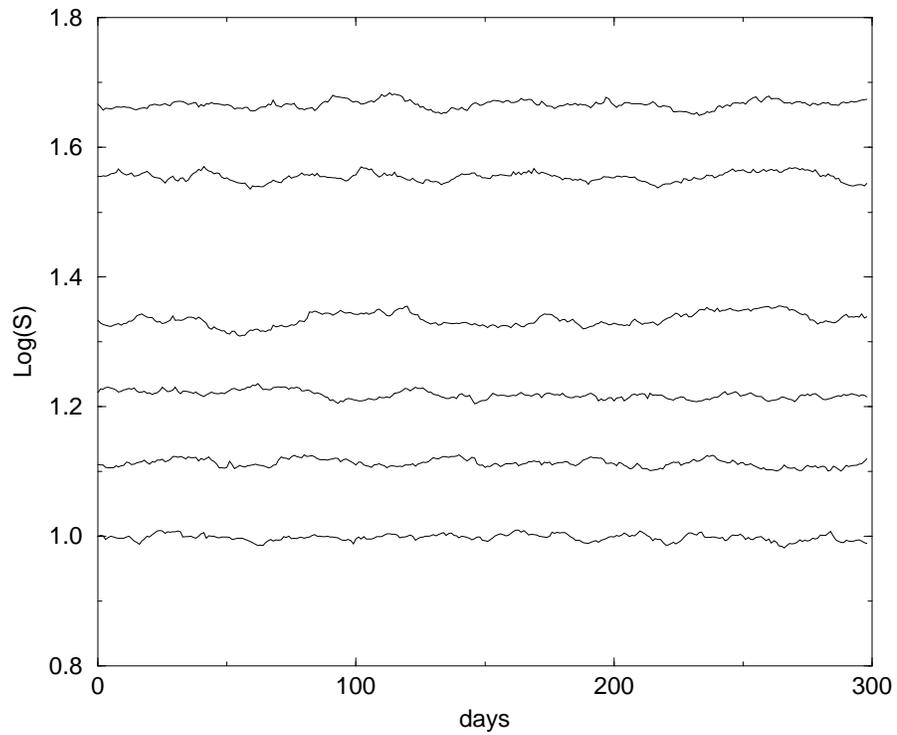}}
\caption{Several thermalized paths for (Black-Scholes) with r=0.05 and 
$\sigma=0.12$ }
\end{figure}

\begin{figure}[th]
  \begin{center}
    \epsfig{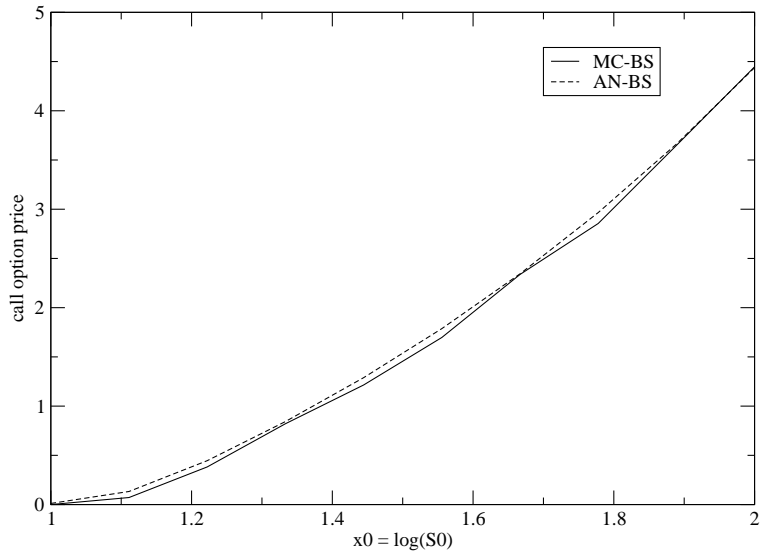}
    \caption{Call option price for strike price 3 versus the logarithm of the initial  value of the stock $x_0=\log(S_0)$. 
      the parameters are fixed as in figs~1. Shown is the analytical result 
      vs the monte carlo result, with a low resolution of 10,000
      configurations}
  \end{center}
\end{figure}
Figs. 1, 2 and 3, 
illustrate some simple results obtained by the monte carlo mehtod. 
For illustrative purposes, we show the behaviour of the 
Black-Scholes model. Fig.~1 shows a typical thermalized path, generated from a given initial value x (at current time $t=\tau$) assuming a maturity of 300 days, while in Fig.~2 we have plotted several path for different starting 
values x of the stock at current time $\tau$. We have chosen an interest 
rate $r=0.05$ and a 12 percent volatility $\sigma$.
Finally, in Fig.~3 we compare the analytical and the numerical 
evaluation of the Black-Scholes option price with a low resolution for 
(\ref{FK}), in order 
to separate the two curves, which otherwise would 
overlap completely, in order to illustrate the convergence of the 
Metropolis algorithm. 

Barrier options can be analized similarly, equivalently, by this method 
or by the Langevin method. We show in Fig.~4 the evaluation of the price of the 
option using the Langevin method in the presence of a 
step potential sitting at a value of the stock price given by $x_0=\log(S_o)$, 
with $S_0=100$. Compared to the Black-Scholes now the price has been 
discounted. 

Applications of the method to the determination of various pricing kernels 
is underway. More details will be given elsewhere \cite{BCM}.

\begin{figure}[th]
\centerline{\includegraphics[angle=0.,width=.9\textwidth]{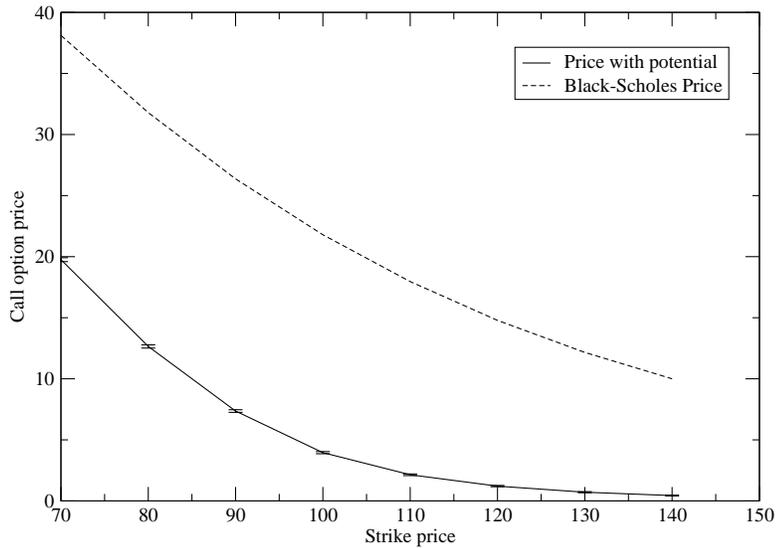}}
\caption{Plot of the option price versus the stock price obtained by a 
Langevin simulation of the path integral, with a potential step}
\end{figure}

\section{Acknowledgements} 
C.C. thanks R. Parwani for hospitality; the National Univ. of Singapore, 
University Scholars Program, for financial support and L. Cosmai 
for previous discussions.


\begin{thebibliography}{0}
\bibitem{BCM} B.E. Baaquie, C. Corian\`{o} and S. Marakani, to appear.
\bibitem{BEB} B.E. Baaquie, {\it Quantum Finance}, to be published.
\bibitem{MG} R. C. Merton, Bell Journal of Economics and Management Science 
{\bf 4} (Spring 1973), 141; M. Garman, {\it A General theory 
of Asset Valuation under Diffusion State Processes.} Working Paper No 50, 
Univ. of California, Berkeley, 1976. 
\bibitem{epstein} J. Epstein and R. Axtell, 
{\it Growing Artificial Societies: Social Science from the Bottom up}, Brookings, MIT Press, 1996. See also the link:{\it www.swarm.org} (Santa Fe Istitute, 
New Mexico). 
\bibitem{rajesh} R. Parwani, {\bf physics/0201055} and links therein. 
\bibitem{BM} B. E. Baaquie, L. C. Kwek and S. Marakani, {\bf cond-mat/008327}
\bibitem{NM}G. Montagna, O. Nicrosini and N. Moreni,
{\it Physica} {\bf A310} (2002) 450; 
G. Montagna, O. Nicrosini {\it Eur. Phys. J.} {\bf B27} 
(2002) 249.

\end{thebibliography}
\end{document}